\documentclass[sigconf, screen, authorversion, nonacm]{acmart}

\definecolor{minor}{HTML}{FF0000}
\usepackage{xcolor}

\usepackage{graphicx}
\usepackage{csquotes}
\usepackage{makecell}
\usepackage{subcaption}
\usepackage{svg}
\usepackage{enumitem}
\usepackage{hyperref}

\AtBeginDocument{%
  \providecommand\BibTeX{{%
    \normalfont B\kern-0.5em{\scshape i\kern-0.25em b}\kern-0.8em\TeX}}}
 
\setcopyright{acmcopyright}
\copyrightyear{2018}
\acmYear{2018}
\acmDOI{10.1145/1122445.1122456}

\acmConference[Woodstock '18]{Woodstock '18: ACM Symposium on Neural
  Gaze Detection}{June 03--05, 2018}{Woodstock, NY}
\acmBooktitle{Woodstock '18: ACM Symposium on Neural Gaze Detection,
  June 03--05, 2018, Woodstock, NY}
\acmPrice{15.00}
\acmISBN{978-1-4503-XXXX-X/18/06}

\begin{document}

%%
%% The "title" command has an optional parameter,
%% allowing the author to define a "short title" to be used in page headers.
\title[Comparing How a Chatbot References User Utterances from Previous Chatting Sessions]{Comparing How a Chatbot References User Utterances from Previous Chatting Sessions:}
\subtitle{An Investigation of Users’ Privacy Concerns and Perceptions}

%Do users give higher quality responses when a health chatbot remembers their previous interactions?

%%
%% The "author" command and its associated commands are used to define
%% the authors and their affiliations.
%% Of note is the shared affiliation of the first two authors, and the
%% "authornote" and "authornotemark" commands
%% used to denote shared contribution to the research.
\author{Samuel Rhys Cox}
%\email{\href{mailto:samcox@comp.nus.edu.sg}{samcox@comp.nus.edu.sg}}
\email{samcox@comp.nus.edu.sg}
\orcid{0000-0002-4558-6610}
\affiliation{%
  \institution{National University of Singapore}
  \country{Singapore}
  }

\author{Yi-Chieh Lee}
\email{yclee@nus.edu.sg}
\orcid{0000-0002-5484-6066}
\affiliation{\institution{National University of Singapore}
  \country{Singapore}
  }

\author{Wei Tsang Ooi}
\email{ooiwt@comp.nus.edu.sg}
\orcid{0000-0001-8994-1736}
\affiliation{\institution{National University of Singapore}
  \country{Singapore}
  }

%%
%% The abstract is a short summary of the work to be presented in the
%% article.
\begin{abstract}
%Does a chatbot referencing previous conversations enhance user engagement or infringe on privacy?
Chatbots are capable of remembering and referencing previous conversations, but does this enhance user engagement or infringe on privacy?
%Chatbots can reference previous conversations, but would this create a tension between users feeling engaged with and feelings of privacy violations?
To explore this trade-off, we investigated the format of how a chatbot references previous conversations with a user and its effects on a user’s perceptions and privacy concerns.
%To explore this trade-off, we compared the effect of 3 different formats a chatbot could use when referencing a user's previous utterances.
In a three-week longitudinal between-subjects study, 169 participants talked about their dental flossing habits to a chatbot that either, (1-None): did not explicitly reference previous user utterances, (2-Verbatim): referenced previous utterances verbatim, or (3-Paraphrase): used paraphrases to reference previous utterances.
Participants perceived Verbatim and Paraphrase chatbots as more intelligent and engaging. However, the Verbatim chatbot also raised privacy concerns with participants.
To gain insights as to why people prefer certain conditions or had privacy concerns, we conducted semi-structured interviews with 15 participants. 
%Upon study completion, we conducted semi-structured interviews with 15 participants to gain insights such as to why people prefer certain conditions or had privacy concerns.
%These findings can help designers choose an appropriate form of referencing previous user utterances and inform in the design of longitudinal dialogue scripting.
We discuss implications from our findings that can help designers choose an appropriate format to reference previous user utterances and inform in the design of longitudinal dialogue scripting.
%We discuss implications for the design and research of voice interactions for navigating instructional videos while performing complex tasks.

\end{abstract}

%%
%% The code below is generated by the tool at http://dl.acm.org/ccs.cfm.
%% Please copy and paste the code instead of the example below.
%%


%%
%% Keywords. The author(s) should pick words that accurately describe
%% the work being presented. Separate the keywords with commas.

\maketitle

%We want to investigate how to improve the quality of user responses given to a health chatbot.
%This is important as more high quality user utterances will allow the chatbot to better understand the user and in turn provide better levels of service (such as providing the user with health guidance).

\section{Introduction}

Advances in language models are leading to chatbot interactions that can persist across multiple sessions, and refer back to previous user utterances \cite{xu2021beyond,xu2022long,park2023generative,bae2022keep}. 
This use of long-term memory can help maintain relationships and build rapport \cite{fivush1996remembering,bluck2003autobiographical,brewer2017remember}, and can improve user experience in chatbot interactions such as in open-domain conversations \cite{xu2022long,xu2021beyond} or discussions of personal health and wellness \cite{bae2022keep,jo2023understanding,holmes2018weightmentor}.
%This use of long-term memory could be used to improve user experience, such as in open-domain conversations \cite{xu2022long,xu2021beyond} or discussions of personal health and wellness \cite{bae2022keep,jo2023understanding,holmes2018weightmentor}.
%or added rapport building and personalisation. 
Additionally, by giving more relevant responses \cite{schuetzler2018investigation,chaves2021should} or explicitly referencing past user utterances \cite{jo2023understanding,rourke1999assessing,lander2015building,kreijns2022social}, a chatbot could increase its social presence: the feeling that it is present in the conversation \cite{oh2018systematic,biocca2002defining,rourke1999assessing}.
%Additionally, by explicitly referencing past user utterances, a chatbot could increase the feeling that it is present in the conversation \cite{jo2023understanding,rourke1999assessing,schuetzler2018investigation}, known as the social presence \cite{oh2018systematic,biocca2002defining,rourke1999assessing} of the chatbot. 
%On from this, health chatbot designers may wish to reference previous conversation details to improve perceived empathy.
%However, previous work has been inconsistent as to whether interviewers should exhibit more or less social presence in order to make interviewees feel comfortable when discussing their information \cite{ng2020simulating,schuetzler2018influence,chen2021you,xiao2020tell,tsaihuman, hagens2023trustworthy}, with some work finding added social presence beneficial \cite{tsaihuman,xiao2020tell}, and others finding social presence detrimental \cite{schuetzler2018influence,chen2021you,ng2020simulating}.

While this could prove beneficial to users and improve user perceptions of the chatbot, it could also lead to feelings of privacy violations. 
This phenomenon is known as the Personalisation Privacy Paradox \cite{awadpersonalization}, where there is a tension between collecting more user data to provide personalised services, and a user's feeling of intrusiveness leading to unwillingness to share their personal information.
%For example, users of mHealth services have reported how concerns over use of their personal data can negatively impact their satisfaction and adoption of services \cite{guo2016privacy}.
%[Are there citations about concerns within chatbot studies? I.e., people find it creepy that a chatbot remembers what they said etc.]
This trade-off could particularly be an issue when people are discussing their sensitive information \cite{gomez2023sensitive}\cite[Art.9]{GDPR}. For example, people may be less willing to disclose socially undesirable behaviours due to embarrassment \cite{tsaihuman}, and users of mHealth services have reported that concerns over use of their personal data can negatively impact service adoption and satisfaction \cite{guo2016privacy}. 

The Personalization Privacy Paradox may hold additional uncertainty when chatbot designers need to choose an appropriate referencing format given the range of styles available to them \cite{zhang2018making,el2021automatic,Cappelen1997-CAPVOQ,wilson_sperber_2000}\footnote{See \cite{wilson_sperber_2000} for a summary of the four main types of quotation from linguistics literature.}.
To explore this paradox, we investigated the level of social presence used when a chatbot references a user's utterances, and its effect on both how privacy violating, and positively (e.g., intelligent, engaging) users perceived the chatbot.
Specifically, we compared 3 referencing formats from low social presence (not explicitly referencing user utterances) to higher social presence (referencing user utterances either verbatim or via paraphrases). 
We conducted a between-subjects longitudinal study ($N = 169$) where participants talked to a chatbot about their dental flossing once a week for three weeks. Participants talked to a chatbot that either: (1-None) did not explicitly reference their previous week's utterances, (2-Verbatim) referenced their previous week's utterances verbatim (e.g.., ``\textit{Last week you said ``My teeth sometimes hurt when I floss''}''), or (3-Paraphrase) referenced their previous week's utterances using paraphrases (e.g., ``\textit{Last week you said that your teeth hurt}'').
Users found chatbots that explicitly referenced their past utterances more intelligent and engaging. 
However, explicitly referencing a user’s past utterances also lead to increased feelings of privacy violations.
To gain further insights as to \textit{why} users might have perceived chatbot referencing formats differently, we conducted semi-structured interviews ($N=15$). 
%In addition, we were also interested in \textit{why} users might perceive chatbot referencing formats differently. Therefore, we conducted semi-structured interviews (N = 15) to gain further insights.
%
%Veratim preferred by those who do not want to lose nuance or prefer to be held accountable to their past utterances, and Paraphrase preferred by those who empasise its more human-like and natural interaction.
Finally, we discuss implications and provide recommendations for chatbot designers when scripting interactions that reference user utterances.

\begin{figure*}[h]
    \centering
    \includegraphics[width=1\textwidth]{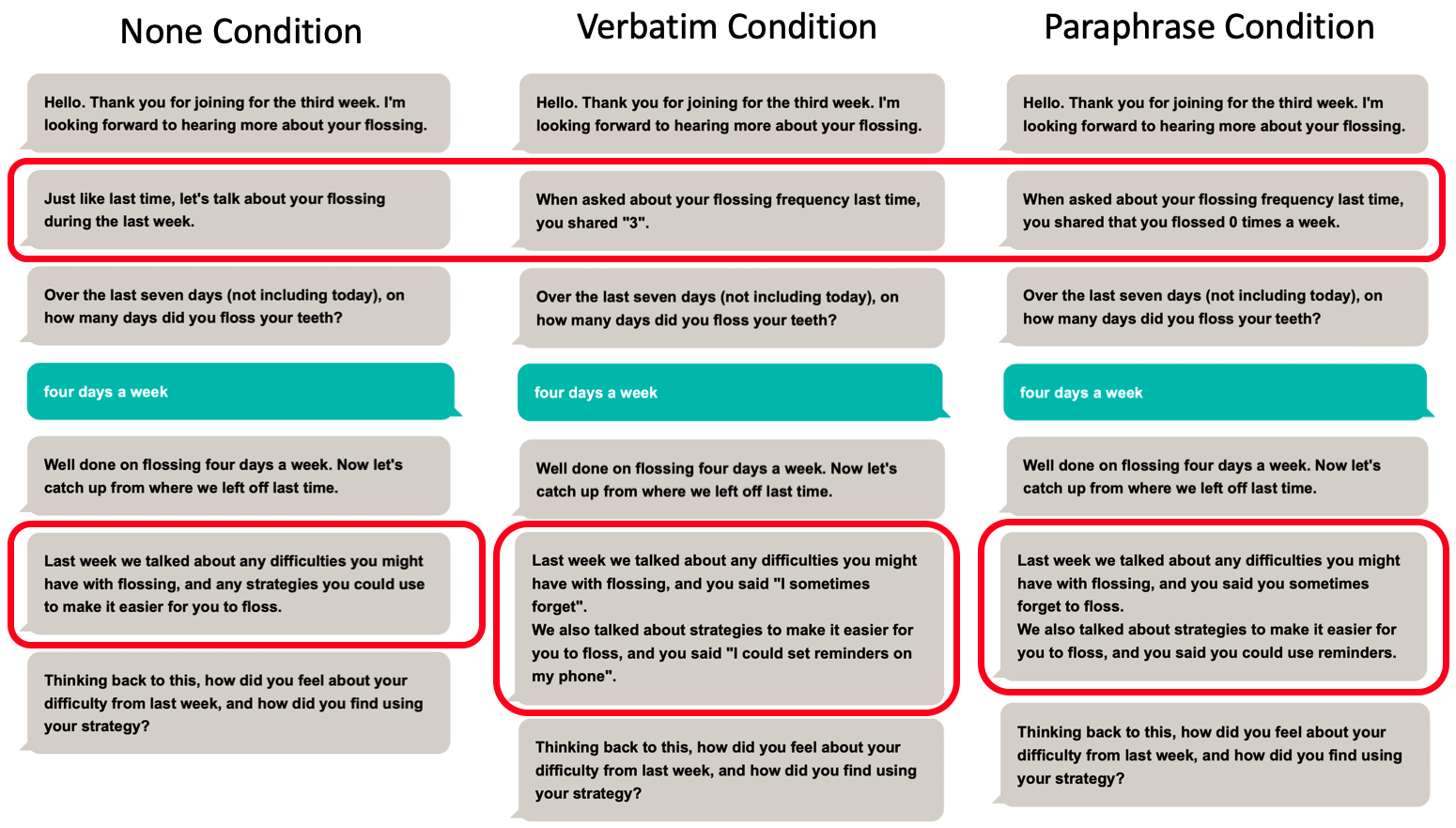}
    \caption{Extracts from week 3 of the study showing the 3 levels of chatbot referencing format.
    Grey bubbles are chatbot utterances, and teal bubbles are user utterances (as seen by participants). 
    Differences in referencing format are circled in red.}
    \label{fig:conditions}
\end{figure*}

\section{Related Work}

%We aim to investigate the effect of a chatbot referencing a user's utterances via different referencing formats.

\subsection{Increasing Social Presence and User Perceptions of Chatbots}
\label{sec:related-social}

As a chatbot uses more conversational referencing formats, it could be said to possess higher levels of social presence \cite{jo2023understanding,rourke1999assessing,lander2015building,kreijns2022social} (i.e., the feeling of being with an interlocutor \cite{oh2018systematic,biocca2002defining,rourke1999assessing}).
%In chatbot studies, the user's feeling of ``being with'' another person when talking to the chatbot has been an objective across multiple studies [CITE]. This feeling of an interlocutor being more human-like and present in a conversation is what is know as the ``social presence'' of said interlocutor \cite{oh2018systematic,biocca2002defining}.
Previous work has investigated the user's feeling of ``being with'' a more present, engaging and human-like chatbot \cite{zheng2022ux,chaves2021should}, 
%In chatbot studies, the user's feeling of ``being with'' a more present, engaging and human-like interlocutor has been an objective across multiple studies \cite{zheng2022ux}, 
%This feeling of an interlocutor being more human-like and present in a conversation is what is know as the ``social presence'' of said interlocutor \cite{oh2018systematic,biocca2002defining}.
and increasing these feelings has been found to have benefits such as improved trust \cite{hagens2023trustworthy,zhou2019trusting} and desire to engage with \cite{pauw2022avatar,lee2020hear, bickmore2000weather,bickmore2001relational} chatbots.
%and increasing these feelings has been found to have multiple benefits such as increasing the positive perceptions of chatbots and increasing the desire for users to engage with chatbots \cite{pauw2022avatar, lee2020hear, zhou2019trusting,bickmore2000weather,bickmore2001relational}.
%Furthermore, chatbots have been found to benefit from various human-like qualities such as empathy \cite{lin_caire_2020}, listening \cite{xiao2020tell}, differing conversational styles and personas \cite{liu2018should,volkel2022user,cox2022does}, and adopting favourable politeness strategies \cite{brown_politeness_1987}.
Furthermore, chatbots have been found to benefit from various human-like qualities such as empathy \cite{lin_caire_2020,liu2018should}, listening \cite{xiao2020tell}, and differing conversational styles \cite{cox2022does,chaves2019s,yang2021effect}, personas \cite{volkel2022user,roy2021users} or politeness strategies \cite{brown_politeness_1987,ouchi2019should,miyamoto2017improving}.
%Increasing the social presence of chatbots has thereby attempted to increase the positive perceptions of chatbots [CITE] and desire for users to engage with chatbots.
%Disclose more to high social presence

Previous studies have also found benefit in interviewers that have higher levels of social presence.
For example, Xiao et al. found that people give higher quality responses to chatbots that use a battery of AI-driven techniques such as using more relevant responses to users \cite{xiao2020tell}; 
Tsai et al. found that users were more likely to disclose embarrassing behaviours related to their sexual health to a human compared to a chatbot \cite{tsaihuman}; 
and multiple studies have found that chatbots that self-disclose information lead to mutual disclosure from users and improved feelings of trust \cite{lee2020designing,adam_onboarding_2019,moon2000intimate,saffarizadeh2017conversational}.
%They measured the quality of responses given by users, but answers sought were not health related, and no single technique can be identified as contributing most to improving response quality (as multiple techniques were used at once).
%Among non-embodied text-based chatbots, increasing the level of a chatbot's self-disclosure has been shown to improve user perceptions of chatbots and improve levels of a user's self-disclosure.
%Disclose more to low social presence:

%They also found that people were less likely to disclose to chatbots which use more relevant responses to user utterances during a small-talk session before asking (non-differentiated) health questions. 
% et al. \cite{x} found that people disclosed less information to human interviewers compared to a electronic survey. They explained this result due to the difference in social presence between humans and surveys, with higher levels of social presence of human interviewers leading users to feel embarrassed or judged when sharing socially undesirable behaviours.

More specifically to our study of chatbot referencing format, previous work has found benefit in chatbots that remember and reference details from previous interactions \cite{jain2018evaluating, portela2017new, chen2021you,medhi2017you,zamora2017m}.
For example, Jain et al. found chatbots that reference details from previous conversations lead to increased feelings of empathy \cite{jain2018evaluating},
%Additionally, chatbots which remember and reference details from previous interactions have led to perceived benefits such as higher levels of empathy \cite{jain2018evaluating}.
and Portela and Granell-Canut reported that participants perceived a chatbot to have higher levels of affection when it remembered previous user utterances or the user's name \cite{portela2017new}.
%Beyond this, we are interested in the effect on positive user perceptions caused by the format used when a chatbot references a user's previous utterances. 
Beyond this, we are interested in the effect on positive user perceptions caused by the format used by a chatbot when referencing a user’s previous utterances.
This gives us our first research question of:

\begin{itemize}
    %\item RQ1: How does the format a chatbot references a user's previous utterances impact a user's:
    %\item \textbf{RQ1:} How does the format of a chatbot's references to a user's previous utterances impact:
    \item \textbf{RQ1:} How does chatbot referencing format (None, Verbatim, Paraphrase) impact:
    \begin{enumerate}[label=(\alph*)]
        \item desire to continue using the chatbot?
        \item perceived chatbot engagement?
        \item perceived chatbot intelligence?
    \end{enumerate}
\end{itemize}

%Chatbots have been used to help users over multiple interactions either to track their personal behaviour, help with mental health \cite{volkel2022user},...
%V\"{o}lkel et al. \cite{volkel2022user} investigated user's interacting with a different personalities of chatbots once a day for four days, and found that people preferred interacting with a chatbot with an extraverted persona.

\subsection{Privacy Concerns Among Chatbot Users}

However, while Section \ref{sec:related-social} outlines the benefits of increased social presence, it could also lead to increased feelings of privacy concerns \cite{xu2008examining} amongst chatbot users \cite{ischen2019privacy,chen2021you}.
%, and this could exacerbated when people are discussing their personal health information \cite{tsaihuman,guo2016privacy}. 
%Following this, some previous work has found that increasing social presence of interviewers can come at the detriment of user trust and privacy concerns.
%In contrast of findings in Section \ref{sec:related-social}, some previous work has found less benefit in interviewers with higher social presence.
%Schuetzler et al. \cite{schuetzler2018influence} found that people disclose lesser amounts of sensitive information to a human than to a chatbot.
For example, Schuetzler et al. \cite{schuetzler2018influence} found that people were less likely to disclose to chatbots that use more relevant responses to user utterances during a small-talk session before asking (non-differentiated) health questions.
Ng et al. showed participants two hypothetical financial chatbots (one human-like and one factual) and found that, while the human-like chatbot scored higher social presence, participants were more likely to share information with the factual chatbot \cite{ng2020simulating}.
%For example, people may be less willing to disclose socially undesirable behaviours due to embarrassment \cite{tsaihuman}, and users of mHealth services have reported how concerns over use of their personal data can negatively impact their satisfaction and adoption of services \cite{guo2016privacy}. 
%This is particularly a concern where people are discussing their personal health information where they may be less willing to disclose socially undesirable behaviours due to embarrassment \cite{tsaihuman}.
%(For a covid-19 survey chatbot), participants found the chatbot more intrusive and rated that they were less likely to comply with advice when it referenced their previous (yes/no) responses \cite{chen2021you}.
Bae et al. \cite{bae2023friendly} found that people trusted a robot-like chatbot more than a human-like chatbot when discussing positive experiences.
More analogous to our study's aim, Chen et al. investigated the perceived invasiveness of a chatbot that referenced participants' personal information (name, presence of heart disease and hand-washing frequency) \cite{chen2021you}.
%Chen et al. asked people to discuss Covid-19 symptoms and behaviours by answering yes/no questions (e.g., ``\textit{Do you have a fever or are you feeling feverish?}'') and two open-text questions (e.g., ``\textit{How often do you wash your hands in a day?}''), while talking to a chatbot that either referenced previous conversations or not \cite{chen2021you}.
While some findings indicated that people found chatbots more invasive when referencing their information, this was contrasted with a null finding once the user's perceived identity of the chatbot (human or chatbot) was taken into account.
%Participants found the chatbot more intrusive and rated that they were less likely to comply with advice when it referenced previous conversations.
%However, Chen et al. did not investigate the effect on the quality of user utterances, did not ask users open-ended questions about \textit{why} they do not adhere to a healthy behaviour, and did not measure for behaviour change.
Building on these previous findings and conflicting results gives us our second research question of:

\begin{itemize}
    %\item \textbf{RQ2:} How does the format of a chatbot's references to a user's previous utterances impact the user's feelings of privacy violations?
    \item \textbf{RQ2:} How does the chatbot referencing format (None, Verbatim, Paraphrase) impact the user's feelings of privacy violations?
\end{itemize}

%We were interested to explore this, as conflicting findings could result in multiple findings.
We were interested to explore RQ2, as conflicting previous work indicates potential contradictory and uncertain findings. 
That is to say, by referencing user utterances in different formats, it could become more apparent to the user that the chatbot is storing or manipulating their personal information, and thereby heighten privacy concerns.
Alternatively, users could appreciate the increased levels of social presence and personalisation.
Additionally, by referencing user utterances verbatim, the chatbot could either make data storage more apparent and therefore privacy-violating to users, or it could be seen as more transparent about storing the user's data without manipulation (and by showing less advanced AI capabilities, users may perceive the chatbot more favourably by generating a metaphor of a chatbot which is less capable \cite{khadpe2020conceptual}).
%Additionally, by referencing user utterances verbatim, the chatbot could either be seen as more transparent about storing the user's data without manipulation (and by showing less advanced AI capabilities, users may perceive the chatbot more favourably by generating a metaphor of a chatbot which is less capable \cite{khadpe2020conceptual}), or it could make data storage more apparent and therefore privacy violating to users. 
Similarly, by paraphrasing user utterances the chatbot could be seen as invasive (by storing and manipulating user data), or create greater feelings of engagement with the user.
Finally, by not explicitly referencing user utterances, the chatbot could be seen as less privacy violating, but also potentially less engaging.

\section{User Study}

This study investigates the effect of a chatbot remembering (and incorporating into conversation dialogue) user utterances from a previous chatting session. 
%For this, we conducted a longitudinal between-subjects experiment recruiting participants from university recruitment pages, where participants talked to a chatbot about their dental flossing once a week for three consecutive weeks.
For this, we conducted a longitudinal between-subjects experiment where participants talked to a chatbot about their dental flossing once a week for three consecutive weeks\footnote{Ethics approval received from our institutional IRB prior to study commencement.}.
Our chatbot had an independent variable of \textbf{Chatbot Referencing Format} (3 levels) which affected whether the chatbot \textit{explicitly} referenced (the previous week's) user utterances, and the format used when referencing utterances (see Figure \ref{fig:conditions} for examples of referencing format). 
The levels of \textbf{Chatbot Referencing Format}\footnote{Literature may refer to referencing formats using various terminology. In our case, verbatim is analogous to extractive summarisation \cite{zhang2018making,el2021automatic} or direct quotation \cite{wilson_sperber_2000,Cappelen1997-CAPVOQ}, and paraphrase is analogous to abstractive summarisation \cite{zhang2018making,el2021automatic} or indirect quotation \cite{wilson_sperber_2000,Cappelen1997-CAPVOQ}.} are:

%We will launch a longitudinal study on Amazon Mechanical Turk (AMT) where participants will discuss their dental flossing behaviour. We aim to investigate the effect of a health chatbot remembering (and incorporating into conversation dialogue) previous user utterances on user perceptions, user health behaviour, and the quality of subsequent user utterances.
%To this end, we will conduct a between-subjects experiment with the independent variable of \textbf{chatbot individuation level} (3 levels):
\begin{itemize}
    \item \textbf{None} (control group): Chatbot did not explicitly incorporate previous user utterances into subsequent conversations, and instead referenced previous discussions at a high-level.
    \item \textbf{Verbatim}: Chatbot incorporated previous user utterances verbatim into subsequent chatbot utterances.
    \item \textbf{Paraphrase}: Chatbot incorporated paraphrased versions of user utterances into subsequent chatbot utterances. 
    %Using data from Chapter \ref{ch:FormalOrCasual}, the paraphrased responses will use topic modelling to recognise the intent of user utterances, and give a response relevant to a user's barriers to flossing (similarly to the method used in \cite{xiao2020if}).
    %We will recognise the intent of user utterances using labelled user utterances from a previous study where users discussed their barriers to flossing with a chatbot. Utterances are labelled for the barrier identified by the user and the strategy to overcome said barrier. Intent recognition can be done via free services online such as IBM Watson or BigML.
\end{itemize}

%Literature may refer to referencing formats using various terminology. In our case, verbatim is analogous to extractive summarisation \cite{zhang2018making,el2021automatic} or direct quotation \cite{wilson_sperber_2000}, and paraphrases are more analogous to abstractive summarisation \cite{zhang2018making,el2021automatic} or indirect quotation \cite{wilson_sperber_2000}.

%We measure the effect of these three conditions on user perceptions and behaviour. Additionally, we conducted semi-structured interviews with participants who completed all three weeks of the study to gain additional insights (see Section \ref{sec:interviews}).

\iffalse
\begin{figure*}[h]
    \centering
    \includegraphics[width=0.75\textwidth]{Figures/Individuation Longitudinal Study Diagram.png}
    \caption{Experiment procedure. Participants were divided into 3 conditions, and talked to a chatbot once a week for 3 weeks about their dental flossing habits and beliefs.}
    \label{fig:individuatation-longitudinal}
\end{figure*}
\fi

\begin{figure*}[h]
    \centering
    \includegraphics[width=1\textwidth]{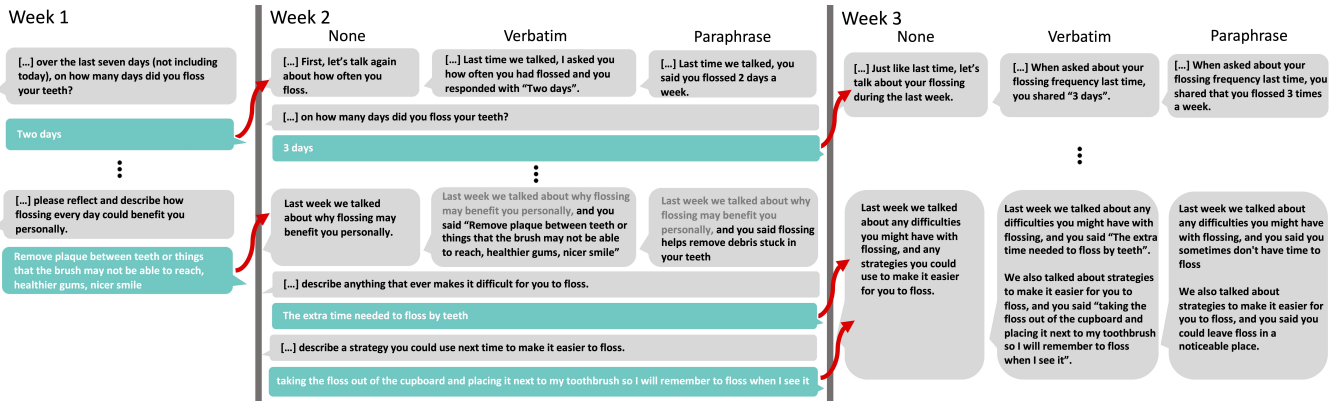}
    \caption{User utterances and their potential None, Verbatim and Paraphrase chatbot responses across all 3 weeks. Grey bubbles are chatbot utterances, and teal bubbles are user utterances. Red arrows show where a user utterance would be referenced (by Verbatim and Paraphrase) in the following week.}
    \label{fig:individuatation-longitudinal}
\end{figure*}

%We chose dental flossing as our domain because it is a health-related behaviour that can be tracked and discussed longitudinally.
%Therefore, the chatbot can elicit and reference user utterances related to a user's health behaviour, beliefs, barriers and strategies to overcome barriers.
%When the chatbot references a user's previous utterances, it is reminding them of their previous health behaviour, beliefs and goals.

%Dental flossing is an activity that can benefit from both diary keeping \cite{suresh2012exploratory} and brief interventions \cite{gillam2019brief}, which health experts recommend daily adherence to \cite{schuz2006adherence}. 

%Accordingly, we chose dental flossing as the domain of our chatbot as we wanted to investigate a use case where a participant could talk to a chatbot about a healthy behaviour to which health experts recommend regular adherence (as is the case for flossing \cite{schuz2006adherence}).
%Additionally, people can have barriers to dental flossing \cite{aguirre2016identification,buglar2010role}, which was key to our chatbot's script; people are benefitted by brief interventions \cite{gillam2019brief} and daily diary keeping \cite{suresh2012exploratory} - tasks which could be facilitated by chatbots; and flossing benefits one's health and is effective at reducing disease causing bacteria \cite{corby2008treatment} and gum bleeding \cite{graves1989comparative}.

\subsection{Chatbot Script}
\label{sec:script}

The chatbot led a conversation with the user about their dental flossing habits and beliefs. 
We chose dental flossing as we want users to discuss something personal to themselves and (as flossing can benefit from both diary keeping \cite{suresh2012exploratory} and brief interventions \cite{gillam2019brief}) it is appropriate for short weekly personal conversations.
%We chose dental flossing as we want users to discuss something personal to them over the course of the study and, as flossing can benefit from both diary keeping \cite{suresh2012exploratory} and brief interventions \cite{gillam2019brief}, flossing is appropriate for this.
Additionally, dental flossing is an activity that health experts recommend daily adherence to \cite{schuz2006adherence}, and people can have barriers to dental flossing \cite{aguirre2016identification,buglar2010role}, both of which are incorporated into our chatbot's script.

%We chose dental flossing as it is an activity that health experts recommend daily adherence to \cite{schuz2006adherence}, and that can benefit from both diary keeping \cite{suresh2012exploratory} and brief interventions \cite{gillam2019brief}. 
%People can have barriers to dental flossing \cite{aguirre2016identification,buglar2010role}, and can therefore reflect and devise strategies to overcome these.

%Each of the 3 weekly conversations with the chatbot was brief and covered a different set of topics.
%All participants were shown the same In week 1, participants shared their dental flossing beliefs and their perceived benefits of flossing.
%In week 2, the chatbot referenced the user's perceived benefits from flossing in week 1, before participants discussed their barriers to flossing, and strategies they could use to overcome this.
%In week 3, participants discussed their susceptibility to negative consequences from not flossing and risks incurred by not flossing.
%Participants discussed their flossing frequency in all 3 weeks.

The conversations for each of the 3 weeks were as follows (responses elicited by the chatbot were open-ended unless specified otherwise)\footnote{Please see supplementary material for chatbot script for all 3 weeks alongside differentiation between the 3 conditions.}:

%Each of the 3 weekly conversations with the chatbot was brief and covered a different set of topics, and the conversation for each week was as follows (responses elicited by the chatbot were open-ended unless specified otherwise):

\begin{itemize}
    \item \textbf{Week 1:} All participants saw the same script as the chatbot could not yet reference previous week's utterances. Participants shared their dental flossing beliefs \cite{buglar2010role} (7-point Likert), flossing frequency, and perceived benefits of flossing.
    \item \textbf{Week 2:} The chatbot referenced flossing frequency and perceived benefits from Week 1. Participants shared their flossing frequency, barriers to flossing, and strategies to overcome barriers.
    \item \textbf{Week 3:} The chatbot referenced flossing frequency, and barriers and strategies from Week 2. Participants shared their flossing frequency, reflected on their barriers and strategies from the previous week, and shared their perceived susceptibility and perceived risks, before sharing their dental flossing beliefs \cite{buglar2010role} (7-point Likert).
\end{itemize}

%Please see supplementary material for the full chatbot script (including highlights of user utterances that were referenced back to the user in the following week's session).

%Please see supplementary material for chatbot script for all 3 weeks alongside differentiation between the 3 conditions.

%To make chatbot utterances more similar in length for each of the 3 conditions, instead of referencing previous user utterances (as in the Verbatim and Paraphrase conditions), the None condition chatbot gave a high-level summary of the previous week's discussion (without explicitly referencing the user's utterances).

\subsection{Implementation Details}

The chatbot was hosted on Qualtrics, and used JavaScript and HTML to emulate the look and feel of a chatbot.
Microsoft LUIS\footnote{https://www.luis.ai/} was used for both intent recognition (in real-time) and for selecting the most appropriate paraphrase for a given week.

Intent recognition was trained using utterances from \cite{cox2022does} for users' barriers to flossing and strategies to overcome barriers. Training data for other prompts was generated by the research team and by piloting the chatbot until a range of responses could be recognised. Data augmentation (e.g., synonym replacement) was then used to generate additional training data.

%Please see full chatbot script in supplementary material for both the chatbot utterances: (1) with intent recognition, and (2) varied by referencing format.

\subsubsection{Intent Recognition:}
\label{subsec:intent}
We used intent recognition (in all 3 conditions) to recognise the intent of user utterances within a week's session. 
An appropriate response would then be appended to the start of the subsequent chatbot utterance.
For example, the chatbot could deliver ``\textit{Well done on flossing five days a week}'' in response to a user's flossing frequency.

%Across all three experiment conditions, we used intent recognition to append pre-written utterances to the 
%All three experiment conditions used retrieval-based dialogue management system with all chatbot utterances being written by the research team.

%Although the conversation was chatbot led, and followed a set dialogue tree, we used intent recognition to append appropriate and LUIS was used for intent recognition of open-ended responses.

%All three experiment conditions used intent recognition (via Microsoft LUIS \footnote{https://www.luis.ai/}), using a retrieval-based dialogue management system with all chatbot utterances being written by the research team.
%Intent recognition was trained using user utterances from \cite{cox2022does} for recognising users' barriers to flossing and strategies to overcome barriers. Training data for other prompts was generated by the research team and by piloting the chatbot until a range of responses could be recognised by the chatbot. 
%To avoid misclassification of user utterances, user utterances with intent recognition called independent LUIS conversational app instances (i.e., there was a separate chatbot endpoint for users' barrier utterances and users' strategies utterances).

\subsubsection{Delivering Paraphrases:}
%The paraphrases used in the Paraphrase condition were pre-written by the research team.
%To deliver paraphrases of user utterances, we recognised the user's intent, and then assigned the user with a paraphrase (for example, see Table \ref{tab:chatbot-paraphrases} for paraphrases used in Week 2). 
To deliver paraphrases of user utterances, first user intent was recognised via LUIS. Each user intent had a corresponding paraphrase written by the research team, that was then used as the paraphrase in the next chatting session (e.g., for flossing benefits, an intent of ``\textit{prevent gum disease}'', was given the paraphrase ``\textit{flossing helps prevent gum disease}'').
%To deliver paraphrases of user utterances, we recognised the user's intent, and then assigned the user with a paraphrase (for example, see Table \ref{tab:chatbot-paraphrases} for paraphrases used in Week 2). 
While this approach is limited in providing a discrete number of paraphrases and not accounting for multiple intents, it ensured that consistent and coherent paraphrases could be delivered to users.
Example script and paraphrases can be seen in Figures \ref{fig:conditions} and \ref{fig:individuatation-longitudinal}, and a full list of paraphrases and script can be found in supplementary material.

\subsection{Participants}

%We recruited participants using university advertisement boards, and only selected participants who did not fully adhere to daily flossing (similarly to previous intervention studies \cite{kocielnik2017send}).
We recruited participants using university advertisement boards. We only selected participants who did not fully adhere to daily flossing (similarly to previous intervention studies \cite{kocielnik2017send}), 
and all responses were completed remotely and asynchronously.
%All participants completed the study in November and December 2022, and all responses were completed remotely and asynchronously.
Participants were paid S\$2 for the first week's session, S\$2 for the second, and S\$3 for completing the third and final week. Weeks 1 and 2 took on average $\sim$3 minutes, and Week 3 $\sim$5 minutes.

169 participants (mean age 22.7; 64\% female) completed all 3 weeks of the study, with 7 participants completing weeks 1 and 2 only, and 4 participants completing week 1 only. We only include data from participants who completed all 3 weeks (with other participants being paid for their completed time, but excluded from analysis), resulting in 55 None, 58 Verbatim, and 56 Paraphrase.

%The average completion time of week 1 was X, week 2 was Y, and week 3 was Z.

%180 participants completed week 1 of the study, 176 completed week 2, and 169 completed all 3 weeks of the study. 

\subsection{Procedure}

Each week, participants were contacted via email and followed the procedure: (1) Follow Qualtrics link to individual chatbot session. (2-\textit{Week 1 only}) give consent  (participants informed responses are stored and analysed).
(3) Brief instructions recap (i.e., no right/wrong answers, responses in English).
(4) Complete weekly chatting session with chatbot.
(5) Post-test questions (see Section \ref{sec:measures}).

Participants were invited to weeks 2 and 3 seven days after completing the previous week's session, and were given three days to complete these sessions.
Responses were controlled so that only desktop or laptop devices could be used.

\subsection{Measures}
\label{sec:measures}

\subsubsection{Weekly Measures:}
At the end of each week's chatbot session, participants rated their experience on 7-point Likert scales (Strongly Disagree to Strongly Agree), and were asked ``\textit{Do you personally agree or disagree that...}'' for the following measures:
\textbf{Interest to continue chatbot usage:} ``I would want to continue using the chatbot'' \cite{xiao2020if};
\textbf{Chatbot engagement:} ``The chatbot seemed engaged in our discussion'', ``I felt the chatbot was NOT paying attention to what I said'' \cite{shamekhi2018face};
\textbf{Chatbot intelligence:} ``The chatbot was intelligent'', ``The chatbot was competent'' \cite{cuddy2008warmth,cox2022does}.

\iffalse
\begin{itemize}
    \item \textbf{Interest to continue chatbot usage:} ``I would want to continue using the chatbot'' \cite{xiao2020if}.
    \item \textbf{Chatbot engagement:} ``The chatbot seemed engaged in our discussion'', ``I felt the chatbot was NOT paying attention to what I said'' \cite{shamekhi2018face}.
    \item \textbf{Chatbot intelligence:} ``The chatbot was intelligent'', ``The chatbot was competent'' \cite{cuddy2008warmth,cox2022does}.
\end{itemize}
\fi

\subsubsection{Privacy concerns, intrusiveness, and risks:}

To investigate whether chatbot referencing style impacts privacy-related measures, (at the end of \textit{week 3 only}) participants responded to the following 7-point Likert scale questions.
%We also wanted to compare whether chatbot referencing style had an impact on privacy-related measures.
%Therefore, (at the end of week 3 only) we took measures for perceived privacy concerns, privacy intrusion and privacy risks \cite{xu2008examining}. Specifically, on a 7-point Likert scale participants were asked  if they personally agree or disagree to the following questions below.
For \textbf{privacy concerns} (referring to concerns that inhibit users from sharing information \cite{xu2008examining}) measures were: ``I was concerned that the chatbot was collecting too much personal information about me'', ``I was concerned about submitting my information to the chatbot''. For \textbf{privacy intrusiveness} (referring to the unwelcome general encroachment into another’s presence or activities  \cite{xu2008examining}) measures were:
%Privacy intrusion: These involve ``the unwanted general incursion of another’s presence or activities'' \cite{xu2008examining}:
``I feel that as a result of this interaction, information about me is out there that, if used, will invade my privacy'', ``I feel that as a result of this interaction, my privacy has been invaded''. For \textbf{privacy risks} (referring to the uncertainty arising from the possibility of an adverse consequence \cite{xu2008examining}) measures were:
%Privacy risks: These are ``uncertainty resulting from the potential for a negative outcome'':
``Personal information was inappropriately used by the chatbot'', ``Providing the chatbot with my personal information involved many unexpected problems''.

\section{User Study Results}
\label{sec:results}

\begin{figure}[h]
    \centering
    \includegraphics[width=0.5\textwidth]{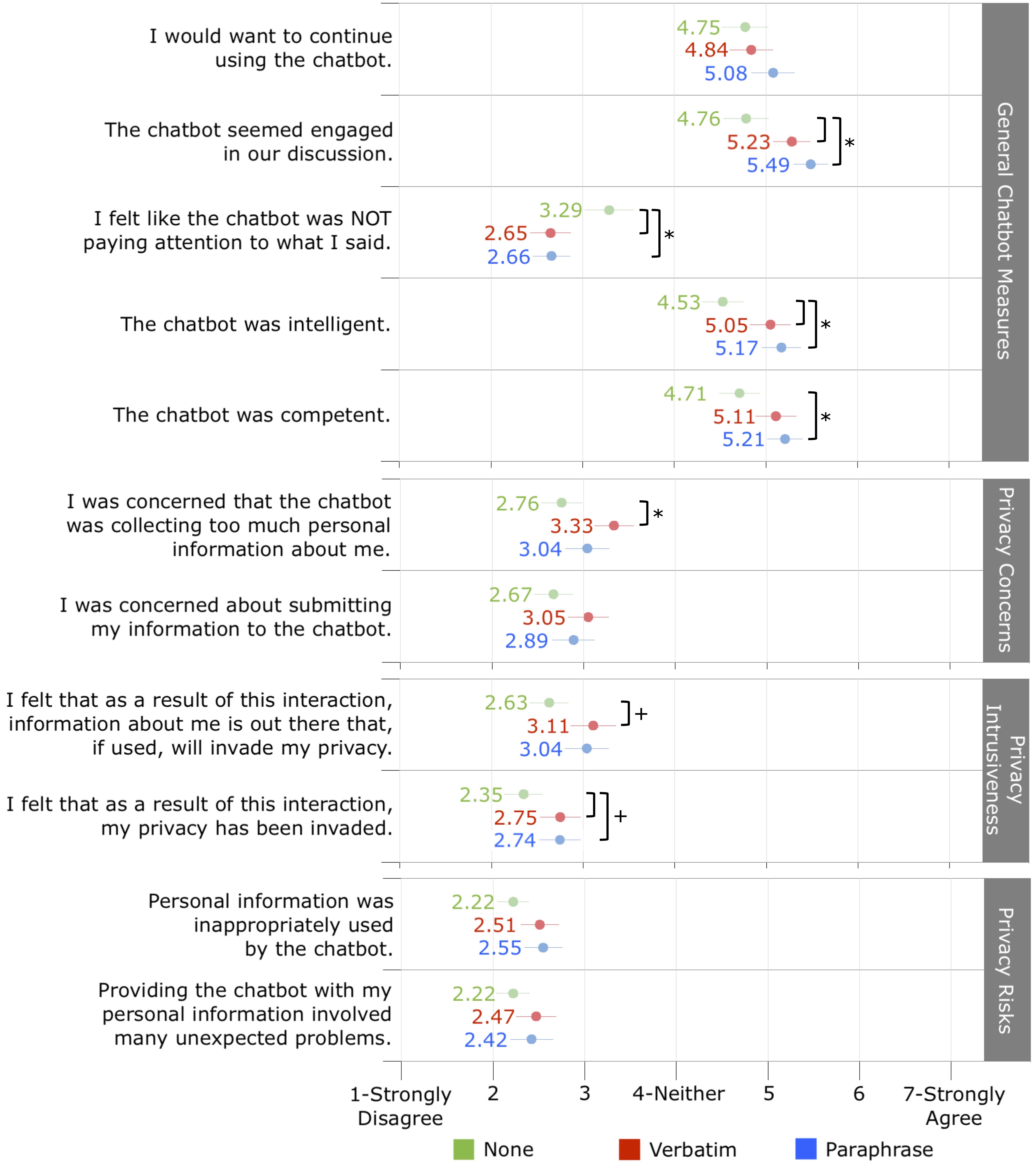}
    \caption{Outcome measures by question asked in the final week of the study. Significance $p<0.05$ indicated by \textbf{$\ast$}, and $p<0.10$ indicated by \textbf{+}.}
    \label{fig:outcomes}
\end{figure}

We fit a linear model on each dependent variable collected from the final week and Chatbot Referencing Format as the fixed effect, and performed post-hoc Student's t-tests to identify specific differences.
We excluded the Likert scale ratings of 8 participants (4 None, 1 Verbatim, 3 Paraphrase) who gave conflicting responses for chatbot engagement (e.g., both rated as Strongly Agree). This left us with 161 responses.
See Figure \ref{fig:outcomes} for summary results. 
In addition, we analysed user responses (response length before and after removing stop words), but found no difference between conditions.
We will now discuss individual findings and their significance.

\subsection{General Chatbot Perceptions}

Measures related to RQ1 are described below. Chatbot referencing format had no direct impact on a user's desire to continue using the chatbot, and there was no significant difference between conditions.

However, participants found the Verbatim and Paraphrase conditions to be more engaging compared to None.
Specifically, for positively perceived engagement, both Paraphrase ($p = 0.0022$) and Verbatim ($p = 0.0444$) were rated more favourably than None.
%Specifically, for positively perceived engagement (``\textit{The agent seemed engaged in our discussion}''), Paraphrase (M = 5.49) scored higher (better) than None (M = 4.76) (p = 0.0022), and Verbatim (M = 5.23) scored higher (better) than None (p = 0.0444).
While Paraphrase scored higher than Verbatim, it was not statistically significant.
%Similarly, for negatively perceived engagement  (``\textit{I felt the agent was NOT paying attention to what I said}''), Paraphrase (M = 2.66) scored lower (better) than None (M = 3.29) (p = 0.0125) and Verbatim (M = 2.65) scored lower (better) than None (p = 0.0159). 
Similarly, for negatively perceived engagement, both Paraphrase ($p = 0.0125$) and Verbatim ($p = 0.0159$) were rated more favourably than None. 
%Both of these results indicate that explicitly referencing a user's previous week's utterances positively impacts the user's perception of engagement and feeling listened to.
These results indicate that explicitly referencing a user's previous week's utterances positively impacts the user's feelings of chatbot engagement, while not explicitly referencing negatively impacts a user's feelings of being listened to.

Participants also found the Verbatim and Paraphrase chatbots to be more intelligent. 
%For perceived intelligence, Paraphrase (M = 5.17) scored higher (better) than None (M = 4.53) (p = 0.0093), and Verbatim (M = 5.05) scored higher (better) than None (p = 0.0301).
For perceived intelligence, both Paraphrase ($p = 0.0093$) and Verbatim ($p = 0.0301$) were rated more favourably than None.
For perceived competence, Paraphrase was rated more favourably than None ($p = 0.0327$).
%Participants found Paraphrase (M = 5.21) to be more competent compared to None (M = 4.71) (p = 0.0327).
%``\textit{The chatbot was competent}''
%Paraphrase scored higher (better) than None (p = 0.0327)
These two results indicate that explicitly referencing a user's previous utterances makes a chatbot appear more intelligent and competent.

\subsection{Privacy Perceptions}
\label{subsec:privacy-concerns}

%Among the two measures of \textbf{privacy concerns}, participants found that the Verbatim chatbot was collecting too much personal information compared to None.
Measures related to RQ2 are described below.
The Verbatim chatbot was found to generate more \textbf{privacy concerns} than None for one of the measures.
Specifically, for ``\textit{I was concerned that the chatbot was collecting too much personal information about me}'', Verbatim scored higher than None ($p = 0.0227$).
%For the measure ``\textit{I was concerned about submitting my information to the chatbot}'', while Verbatim (M = 3.05) scored the highest (worst), it was not statistically significant to None which had the lowest score (M = 2.67).%(p = 0.1130)

For measures of \textbf{privacy intrusiveness}, there were weakly significant differences, that \textit{could} suggest participants found Verbatim or Paraphrase conditions to be more intrusive compared to None.
For the measure: ``\textit{I feel that as a result of this interaction, information about me is out there that, if used, will invade my privacy}'', Verbatim scored highest (worst) and was weakly different to None ($p = 0.0716$).
%For the measure: ``\textit{I feel that as a result of this interaction, my privacy has been invaded}'', Verbatim (M = 2.75) scored highest (worst), and was weakly different to None (M = 2.35) ($p = 0.0677$).
%Similarly, Paraphrase (M = 2.74) was weakly different to None ($p = 0.0866$).
For the measure: ``\textit{I feel that as a result of this interaction, my privacy has been invaded}'', both Verbatim ($p = 0.0677$) and Paraphrase ($p = 0.0866$) scored higher than and were weakly different to None.

For both measures of \textbf{privacy risk}, while Verbatim and Paraphrase trended above None, there were no statistically significant differences between conditions.
%``\textit{Personal information was inappropriately used by the chatbot}''
%No significant difference
%Paraphrase is highest scoring (worst), but not significantly different to None (p = 0.1113)
%``\textit{Providing the chatbot with my personal information involved many unexpected problems}''
%No significant difference

These results indicate that explicitly referencing a user's previous utterances may raise privacy concerns, and that this may be further exacerbated if utterances are referenced verbatim. 
%That Verbatim participants were more concerned that the chatbot was collecting too much information about themselves, may also indicate that by directly quoting a user's utterances, people were more conscious of their data being collected, and therefore had more privacy concerns.
In particular, Verbatim participants were more concerned that the chatbot was collecting too much information about themselves. This may indicate that directly quoting a user's utterances made users more conscious of their data being collected, and therefore increased privacy concerns.

However, it is important to note that all privacy measures averaged below ``\textit{4 - Neither Agree Nor Disagree}'' reflecting that feelings of privacy violations were still low amongst participants. This feeling may be reflected by the domain of the chatbot (dental flossing) which some participants may not find to be a very sensitive topic (discussed further in Section \ref{sec:interview_privacu_unrelated}).

\section{Semi-Structured Interviews}
\label{sec:interviews}

Our quantitative results found Verbatim raised more privacy concerns than None, and there were also trending (but weakly-significant) results to indicate that participants found Verbatim and Paraphrase potentially more intrusive than None (see Section \ref{subsec:privacy-concerns}).
To gain further insights as to \textit{why} people may perceive the chatbot referencing formats differently, we conducted semi-structured interviews.

%Verbatim participants statistically significantly felt concerned that the chatbot collected too much of their personal information compared to None. There were also trending, but weakly-significant, results to indicate that participants found the Verbatim and Paraphrase conditions as potentially more intrusive than None (see Section \ref{subsec:privacy-concerns}). 
%Therefore, to gain further insights as to \textit{why} people may perceive the chatbot referencing formats differently, we conducted semi-structured interviews.

%We conducted semi-structured interviews to investigate why people may prefer different forms of a chatbot referencing their previous utterances. 

\subsection{Participants}
We recruited 5 participants per condition (N = 15; mean age 20.9; 9 female) for remote interviews, all of whom had completed the full 3 weeks of the study. Interviews lasted between 20 and 30 minutes, and participants were reimbursed S\$5 for their time.
%, and each interview was conducted online.

\subsection{Procedure}
First, participants were instructed that there are no right or wrong answers, and consent was sought to record the interview.
Participants then discussed their experience taking part in the study and responded to questions pertaining to perceived effect on dental flossing, privacy violations, chatbot intelligence, chatbot warmth and the participant's perception of their assigned condition.

After these questions, the interviewer concluded by revealing and describing the 3 experiment conditions.
%, before asking the participant to describe their feelings towards each condition, and their preferred referencing format. Participants were asked to think-aloud \cite{charters2003use}, and explain their reasoning towards each experiment condition.
Participants were then asked to think-aloud \cite{charters2003use}, and rank their preference for the conditions while explaining their opinion and reasoning.
See supplementary material for the interview guide used.

\iffalse
For the interview questions participants were asked to describe their:
\begin{enumerate}
    \item flossing habits before, during and after talking to the chatbot,
    \item feelings of comfort in sharing information with the chatbot,
    \item opinions on how the chatbot referenced their past utterances (both how they found their assigned experiment condition, and what an ideal alternative could look like)
    \item opinion on the chatbot's intelligence
    \item opinion on the chatbot's warmth
    \item likelihood to recommend the chatbot, and other feedback (if any).
\end{enumerate}
\fi

\subsection{Findings}

%When discussing their privacy concerns, participants had privacy concerns related to both their general perception of chatbots and of the sensitivity of discussing flossing, and privacy concerns related the the referencing style used by the chatbot.

We will now discuss the findings from our semi-structured interviews. 
We discuss privacy concerns raised by participants (split between those related and not related to referencing format), chatbot intelligence, recall assistance, and chatbot naturalness. Finally, we discuss the last section of the interview where participants saw all 3 conditions and explained their referencing format preferences.

%When participants were discussing their privacy concerns, they expressed apprehensions related both to the referencing style deployed, and more general privacy concerns.

%When discussing their privacy concerns, participants expressed apprehensions about two aspects: their overall impression of chatbots and the sensitivity of discussing flossing, as well as the referencing style employed by the chatbot.

\subsubsection{\textbf{Privacy Concerns (Unrelated to Reference Format):}}
\label{sec:interview_privacu_unrelated}

Similarly to previous findings, the perceived sensitivity of the domain varied among participants, and affected their hesitancy in sharing information \cite{cox2022does,markos2017information}.
Some participants without privacy concerns described dental flossing as a non-sensitive domain, meaning they were not hesitant sharing their information:
\begin{displayquote}
``\textit{this topic isn't something that's very sensitive, so I wasn't particularly concerned about it.}'' - P5(Verbatim)
%``\textit{this topic isn't something that's very sensitive, so I wasn't particularly concerned about it. Yeah, but also the chatbot didn't give any reason for me to feel like my there's an invasion of privacy or whatever}'' - P5(Verbatim)
\end{displayquote}

In contrast, some participants were concerned to share their dental flossing behaviour as they saw it as sensitive information.
P11(Verbatim) raised this concern in addition to hesitancy sharing their information from uncertainty as to who will read their messages:

\begin{displayquote}
``\textit{dental flossing is} [...]\footnote{Due to space limitations, some transcriptions have been shortened to remove word repetitions, or thinking aloud speech before interviewees arrived at a final conclusion.} \textit{a more private, embarrassing.. umm.. thing. So I think differently sometimes, like whether telling the chatbot like how often I floss or whether I managed to achieve my goals} [...] \textit{it does feel a little scary because I'm not very sure who exactly is seeing the information.}'' - P11(Verbatim)
\end{displayquote}
%\begin{displayquote}
%``\textit{Yeah, so I think like-like I said like dental flossing is kind of like a-a bit of a more private, embarrassing.. umm.. thing so I think differently sometimes like whether telling the chatbot like how often I floss or whether I managed to achieve my goals it's not that I wouldn't care. I mean 'cus.. that's partly because it's the instructions, but it does feel a little scary because I'm not very sure who exactly is seeing the information.}'' - P11(Verbatim)
%\end{displayquote}

On from this, participants described feeling embarrassed when sharing health behaviour that they perceived as insufficient:
%thought of as too low:

\begin{displayquote}
``\textit{\textit{I was a bit embarrassed, because they asked me, uh, how many times did I floss over the week, and then I was like ``0''}}'' - P4(Paraphrase)
\end{displayquote}
%\begin{displayquote}
%``\textit{\textit{Well, I mean I was a bit embarrassed, because they asked me, uh, how many times did I floss over the week and then I was like ``0'' because I just didn't.. I just didn't floss. So I was like ``oh shit'' a bit.. A bit embarrassed, but then I realise it's fine. Like it's not... Yeah it's not that big of a deal. So, it's just aside from that, like. It was OK. Like the other things, I didn't mind sharing honestly, yeah.}}'' - P4(Paraphrase)
%\end{displayquote}

Furthering this, P12(Paraphrase) felt embarrassed when their flossing frequency was referenced:

\begin{displayquote}
``\textit{I was like... slightly embarrassed} $<$laughs$>$ \textit{about how I-I never floss at all.} [...] \textit{it made me health-aware about how I wasn't really flossing at all} [...] \textit{I wouldn't really say that made me feel uncomfortable. Just like a little bit embarrassed. A little bit self-aware.}'' - P12(Paraphrase)
\end{displayquote}
%\begin{displayquote}
%``\textit{I guess I was like.. slightly embarrassed <laughs> about how I-I never floss at all. Because... Like, yeah, it made me health-aware about how I wasn't really flossing at all, but that's about it ..and I I still have.. I wouldn't really say that made me feel uncomfortable. Just like a little bit embarrassed. A little bit self-aware.}'' - P12(Paraphrase)
%\end{displayquote}

%One participant described how they considered lying to the chatbot about their flossing behaviour, matching previous findings about socially desirable responding \cite{schuetzler2018influence-2,schuetzler2018influence}, and indicating a potential mirroring of previous results that when a chatbot appears to understand users, they disclose less \cite{schuetzler2018influence}.

Similarly to previous findings on socially desirable responding \cite{schuetzler2018influence-2,schuetzler2018influence}, one participant described how they considered lying to the chatbot about their flossing behaviour:

\begin{displayquote}
``\textit{I felt guilty for like flossing my teeth like once a week. And I was like ``try to lie to them'', but then I was like: ``OK, never mind I won't lie to them''}'' - P2(None)
\end{displayquote}

The expectation of data storage and perceived sensitivity of the task also affected feelings of privacy invasion, with P1(Paraphrase) equating the task and data storage to writing a diary for themselves:
%It felt like a diary to them, so it didn’t feel invasive when referencing their utterances:

\begin{displayquote}
``\textit{it's normal for it to store information.} [...] \textit{It's equivalent as to you writing a diary, so it wasn't really something that was particularly invasive to me.}'' - P1(Paraphrase)
%``\textit{I felt like you know it was just a normal thing like, it was just very casual. Very, very normal. Like it's normal for it to store information. If you're trying to use it as a self-help, a self-help thing, it's equivalent for me. It's equivalent as to your writing a diary, so it wasn't really something that was particularly invasive to me. To me it was just like, ``Oh yeah, it's just a little diary that I write to myself''}'' - P1(Paraphrase)
\end{displayquote}

\subsubsection{\textbf{Privacy Concerns (Related to Reference Format):}}

When discussing privacy concerns, several participants expressed surprise that the chatbot referenced what they said in previous weeks:

\begin{displayquote}
``\textit{at first when they repeated what I said the previous week, then I was like, ``oh shit, they record everything'' but-but it's not that big of a deal, I guess. Like it's alright, it's just dental hygiene}'' - P4(Paraphrase)
\end{displayquote}

Some of this surprise was accounted to participants' (lack of) expectations of chatbot abilities, with interviewees describing their concerns subsiding after the initial exposure to chatbot referencing.

However, some Verbatim participants were negatively surprised.
P11(Verbatim) found it ``unnerving'' that Verbatim remembered what they said, and found sharing flossing embarrassing:

\begin{displayquote}
``\textit{it was like a little unnerving because the chatbot remembered what I said previously.} [...] \textit{I didn't expect it to be that smart} $<$laughs$>$\textit{, so it was a little startling but, because-because we're talking about something like dental flossing so, I guess it was a little embarrassing at first.}'' - P11(Verbatim)
\end{displayquote}

%\begin{displayquote}
%``\textit{I think sometimes it-it, \textbf{it was like a little unnerving because the chatbot remembered what I said previously}. So.. umm.. I-I think it was just a bit.. *pause* I didn't expect it to be there. It's... I-I don't know how to phrase this, but I didn't expect it to be that smart <laughs>, so it was a little a little startling but, because-because we're talking about something like dental flossing so, I guess it was a little embarrassing at first, but after a while I think I found it OK.}'' - P11(Verbatim)
%\end{displayquote}

Conversely, P1(Paraphrase) described how the referencing format did not raise feelings of privacy intrusion as they expected their data to be stored:

\begin{displayquote}
``\textit{I think that the chatbot just took in} [...] \textit{whatever I put in from the last time around, and} [...] \textit{data storage is} [...] \textit{a normal thing of a chatbot for me, so not really is anything that felt very intrusive}'' - P1(Paraphrase)
%``\textit{I think that the chatbot just took in whatever I took whatever I put in from the last time around, and so like I mean, data storage is going to be already.. like it's a normal thing of a chatbot for me, so not really is anything that felt very intrusive}'' - P1(Paraphrase)
\end{displayquote}

%P9(Verbatim) described how they appreciate their utterances being unchanged, as any ``processing'' would have raised privacy concerns:
P9(Verbatim) described appreciating utterances being unchanged, as any ``processing'' would have raised privacy concerns:

\begin{displayquote}
``\textit{I would prefer this over if they were to process my message} [...] \textit{Rather that they just feedback what I have said so my perception would be: ``}[...] \textit{they have just stored my data and then given it back to me'' as compared to them }[...] \textit{processing the background and feeding something else, which I think would have raised a bit more of a privacy issue for me.}'' - P9(Verbatim)
%``\textit{I would prefer this over if they were to process my message and I guess find some key theme [...] Rather that they just feedback what I have said so my perception would be: ``oh, they just have like, they have just stored my data'' and then given it back to me as compared to them, I don't know, processing the background and feeding something else. Yeah, which I think they would have raised a bit more of a privacy issue for me, yeah?”}'' - P9(Verbatim)
\end{displayquote}

When asked what made them hesitant sharing information, some None participants described how the non-explicit referencing format of None made them doubt the engagement of the chatbot and thereby be hesitant in sharing information:

%When asked what made them hesitant sharing information, P7(None) described how the non-explicit referencing format of None made them doubt the engagement of the chatbot and thereby be hesitant in sharing information:

\begin{displayquote}
``\textit{I think the only thing that was uncomfortable was that the chatbot} [...] \textit{didn't really seem to engage in the conversation.} [...] \textit{he made a reference to its own question. He didn't make reference to my answer.} [...] \textit{he just made the whole chat feel a bit disengaged. So like whatever answer I put down doesn't really matter to the chatbot anyway}'' - P7(None)
\end{displayquote}
%\begin{displayquote}
%``\textit{I think the only thing that was uncomfortable was that the chatbot [...] didn't really seem to engage in the conversation. So basically, when you made a reference to a previous interaction, he made a reference to its own question. He didn't make reference to my answer.
%So I think it was that he just made the whole chat but feel a bit disengaged. So like whatever answer I put down doesn't really matter to the chatbot anyway - it just replied the same thing}'' - P7(None)
%\end{displayquote}

%Participants in None also reported being hesitant to share information with the chatbot due to its perceived lack of intelligence. P6(None) mentioned that they attempted to simplify their responses in order to improve the chatbot's understanding:

This went further with some None participants describing simplifying their responses as they did not think the chatbot would understand them otherwise:

\begin{displayquote}
``\textit{probably if anything} [made me hesitant sharing information], \textit{it was maybe like how complex I structured my sentences. So I tried to keep my sentences like as simple as possible so that maybe the chatbot would be..} $<$pause$>$ \textit{easier for the chatbot to recognise the sentence structures}'' - P6(None)
\end{displayquote}

%However, some None participants described that None did not violate their privacy as it did not reference their utterances explicitly:

However, some None participants described lack of privacy concerns due to no explicit references to their utterances:

\begin{displayquote}
``\textit{it was just a series of prompts that doesn't really consider any reference to my own, and so I don't really feel any breach of privacy or something}'' - P7(None)
\end{displayquote}

%\begin{displayquote}
%``\textit{because, uh, the summary is actually more of a very general, but not specific to our response, so I think it's good for users that might have privacy concerns because it does not show that they collected their data from previous times.}'' - P10(None)
%\end{displayquote}

%On from this, unique to None were interviewees describing how they might simplify their responses so that the chatbot would be able to understand them better.

%None participants described that the perceived lack of intelligence in the chatbot made them hesitant sharing information, and that they tried to simplify their responses so that the chatbot would understand them better:

%% Group TAG with other "None simplifying responses" quote:
%P6(None) did not trust in the intelligence of the chatbot, and tried to use more simple language:

\subsubsection{\textbf{Perceived Intelligence and Engagement:}}

%Interviewees generally found the Verbatim and Paraphrase chatbots to be more intelligent.
Interviewees generally viewed Verbatim and Paraphrase chatbots as intelligent.

\begin{displayquote}
``\textit{I was like pretty pleasantly surprised that it like remembered my answers from previous weeks. Yeah, It made me think the chatbot was like a little bit more intelligent.}'' - P12(Paraphrase)
%``\textit{I think I was like pretty pleasantly surprised that it like remembered my answers from previous weeks. Yeah, It made me think the chatbot was like a little bit more intelligent.}'' - P12(Paraphrase)
\end{displayquote}

%\begin{displayquote}
%``\textit{I think it was quite intelligent, because it was able to reference previous things that I said}'' - P11(Verbatim)
%\end{displayquote}

Similarly, Verbatim and Paraphrase participants found referencing their previous utterances made the chatbot feel engaged.

%\begin{displayquote}
%``\textit{I think the referencing does help a lot, 'cus then I'm like: ``oh, so the chatbot remembers what I am talking about and it replies like specifically according to what I replied'', so yeah, I think it was quite engaged.}'' - P11(Verbatim)
%``\textit{Yeah, so I'm going to take like engaged as like.. as like the chatbot listening to me. Yeah, I think it was quite high 'cus.. umm.. yeah.. I think the referencing does help a lot, 'cus then I'm like: "oh, so the chatbot remembers what I am talking about and it replies like specifically according to what I replied", so yeah, I think it was quite engaged. It's not like tossed out like irrelevant information to me.}'' - P11(Verbatim)
%\end{displayquote}

However, some participants thought less of Verbatim with \\P5(Verbatim) stating: ``\textit{it felt like a survey}''. Others disliked Verbatim due to its repetition of their utterance word-for-word:

\begin{displayquote}
``\textit{it’d be good to somehow be able to paraphrase what I've said} [...] \textit{so it wouldn't feel so obvious that it's just copying and pasting what I've said previously}'' - P5(Verbatim)
\end{displayquote}

%\begin{displayquote}
%``\textit{Maybe to make it more personalised, can remove the quotation marks} [...] \textit{maybe change it to like} [...] \textit{``last week, you said that something, something, something''. So, without the quotation marks}'' - P13(Verbatim)
%\end{displayquote}

By contrast, while None participants described the intent recognition as a feature of an intelligent chatbot, they also (due to None's referencing format) questioned the intelligence of the chatbot, with some doubting the chatbot's ability to understand them.

\subsubsection{\textbf{Referencing Format and Recall:}}
\label{sec:referencing_recall}

Participants described how the referencing from both Verbatim and Paraphrase helped them remember what they wrote previously.
%\begin{displayquote}
%``\textit{it was a good reminder for me because sometimes I forget what I wrote.}'' - P3(Paraphrase)
%``\textit{I felt like it was a good reminder for me because sometimes I forget what I wrote, like what I wrote as well. So the fact that the bot told me what I did and I'm like "Oh yeah, I did say that" then I.. I can like.. I remember what I was saying the previous time}'' - P3(Paraphrase)
%\end{displayquote}
Verbatim was preferred by some participants as a more precise reminder of their utterance. 
%The referencing format of Verbatim was described by P15(Verbatim) as:
For example, P15(Verbatim) equated the referencing style to a lecture recap, and valued Verbatim's consistency:

\begin{displayquote}
``\textit{Like in lectures and like videos where there's like a recap or review.} [...] \textit{I wouldn't have remembered what I said to the robot, so it kept like a certain consistency of like the interview}'' - P15(Verbatim)
%``\textit{Like in lectures and like videos where there's like a recap or review. It really helps me, you know, sometimes because I-I do, I do have like a pretty bad memory. So, with all of the stuff that's been going on like it was also an exam week, so I think it was exam yeah? I wouldn't have remembered what I said to the robot, so it-it kept like a certain consistency of like the interview and it made me not repeat like answers or like at least update my answers. Like yeah, I think it's a great. Uh, a great feature in the robot.}'' - P15(Verbatim)
\end{displayquote}

%Specifically, they may desire to be held accountable for their prior statements: 

%distrust the chatbot's ability to accurately paraphrase their words and believe paraphrasing will lose nuance

%or consider Verbatim will better distinguish their own utterances from the chatbot's:

Otherwise, participants appreciated Verbatim as they:
distrust a chatbot's ability to accurately paraphrase their words (and believe paraphrasing will lose nuance); 
want to know their exact utterance so previous conversations are not repeated; 
consider Verbatim will better distinguish their own utterances from the chatbot's; 
%consider that their own utterance is more easily distinguishable from the chatbot's utterance
or may desire to be held accountable to their prior utterances: 

\begin{displayquote}
``\textit{the retrieval by the chatbot to bring back exactly, especially word for word, what I said, kind of reminded me that ``ohh, I kind of agreed to this, to try this strategy'' and yeah to see one week later I actually did carry it out}'' - P9(Verbatim)
\end{displayquote}

By contrast, the None participants found referencing utterances at a high-level negatively impacted recall:

\begin{displayquote}
``\textit{the problem in very generic statements is that} [...] \textit{I kind of like forgot what I've written, and then when they tried to resume conversation, I had no idea what I said.}'' - P2(None)
%``\textit{the problem in very generic statements is that I think by the third time I forgot what I've written. In the second time, I kind of like forgot what I've written, and then when they tried to resume conversation, I had no idea what I said, so I was like "Uhh.. what did I say...?". Yeah, you know, like that kind of feeling so I think it... For me I thought it would have been better in the chatbot could like mention what I said at least... yeah because yeah, because I completely forgot about it then...}'' - P2(None)
\end{displayquote}

\noindent Which led some None participants to suggest the chatbot should reference their previous utterances:

\begin{displayquote}
``\textit{it would have been better if the chatbot could like mention what I said at least}'' - P2(None)
\end{displayquote}

%\begin{displayquote}
%``\textit{if it's more specific, it might allow the user to remember what is their response previously and then give a better response for the following answer as well.}'' - P10(None)
%\end{displayquote}

For example, P10(None) suggested that the chatbot could reference utterances similarly to existing messaging applications:

\begin{displayquote}
``\textit{it can be more like} [...] \textit{in WhatsApp or Telegram you can reply to the message.} [...] \textit{so you can actually see that ``actually the chatbot is referring to this message that I have sent previously'', so it is clearer.}'' - P10(None)
%``\textit{I think it can be more like a real message. Like you know in WhatsApp or Telegram you can reply to the message. Like at the top that. That thing.. so you can actually see that "actually the chatbot is referring to this message that I have sent previously", so it is clearer.}'' - P10(None)
\end{displayquote}

\subsubsection{\textbf{Naturalness of Referencing Format:}}
\label{sec:interview_natural}

Participants described Paraphrase as feeling natural and human-like. 
For example, \\P3(Paraphrase) appreciated that the chatbot did not copy previous utterances word-for-word, and thereby felt more engaging: 
%This went so-far as P3(Paraphrase) describing that they appreciated that the chatbot did not copy their previous utterances word-for-word, and thereby felt more engaging: 

\begin{displayquote}
``\textit{how the bot referenced it feels very natural.} [...] \textit{it didn't copy what I said verbatim. 
So like it felt as if like a friend was just like, ``Oh yeah, I remember you said something about this like last time we met'' so it felt quite natural, and} [...] \textit{I also really like the fact that they did remember} [...] \textit{because then it made me feel like ``OK, at least the bot is listening to what I say. I'm not like shouting into the abyss''}'' - P3(Paraphrase)
%``\textit{I think how the bot referenced it feels very natural. It didn't feel like it copied my word, like it didn't copied what I said, verbatim.  So like it felt as if like a friend was just like, ``Oh yeah, I remember you said something about this like last time we met'' so it felt quite natural, and I think I also really like the fact that they did remember, or they did reference what I said, because then it made me feel like “OK, at least the bot is listening to what I say I'm not like-like shouting into the abyss” if that makes sense}'' - P3(Paraphrase)
\end{displayquote}

Conversely, some Verbatim participants described how quoting verbatim did not feel personable:

\begin{displayquote}
``\textit{I feel like because it's.. it was quoted directly, right? I felt like there wasn't, say, a lot of personal interaction. It felt more like those things... are just coded.}'' - P5(Verbatim)
\end{displayquote}

Expanding on this P5(Verbatim) described how they would prefer it if the chatbot could paraphrase their utterances:

\begin{displayquote}
``\textit{it’d be good to somehow be able to paraphrase what I've said, or to do so without directly quoting? Yeah, so it wouldn't feel so obvious that it's just copying and pasting what I've said previously, yeah?}'' - P5(Verbatim)
\end{displayquote}

Some None participants described the condition as less natural, and suggested that explicitly referencing past utterances would make the chatbot more personable:

\begin{displayquote}
``\textit{if they reference to my difficulties directly, you feel more... personal.}'' - P7(None)
%``\textit{Maybe references that relate to my answer instead. So one of the challenges I think I put was that, uhh... Sometimes it's difficult to floss because I have braces, so it didn't really mention about it at all. So maybe if they reference to my difficulties directly, you feel more... personal.}'' - P7(None)
\end{displayquote}

%\begin{displayquote}
%``\textit{probably if anything, it was maybe like how complex I structured my sentences. So I tried to keep my sentences like as simple as possible so that maybe the chatbot would be.. *pause* easier for the chatbot to recognize the sentence structures or what I was trying to explain.
%Yeah, I think one of the weeks. I.. umm.. I did mention like one of my strategies is for example to put the floss nearby me, but then it seemed like the chatbot didn't really understand what I was saying on that. So, I just tried to keep my answers easier to understand.}'' - P6(None)
%\end{displayquote}

\subsubsection{\textbf{Comparing the 3 Referencing Formats:}}

At the end of the interview, we revealed the 3 referencing formats to participants, and asked them to think-aloud and explain their preference between formats.
This reinforced some of the previous qualitative findings, and also generated opinions from participants of their non-assigned conditions.
When ranking their preference for referencing format, all interviewees put None as their last choice, 5 interviewees put Verbatim as their first choice, and 10 interviewees put Paraphrase as their first choice.

Some user feedback, mirrored that discussed in Sections \ref{sec:interview_privacu_unrelated} to \ref{sec:interview_natural}, with users describing Verbatim as ``creepy'', ``scary'' and ``guilt-tripping'' them, or stating that they appreciate the fidelity to their original utterance; Paraphrase as more natural and human-like; and None as unengaging.
Interestingly, some participants who chose Verbatim as their first preference described that they see chatbots as a tool, and value their own word over that of a robot. In contrast, those who favoured Paraphrase described seeing a chatbot as a conversational partner that they wish to be more human-like.

\iffalse
\section{Summary of Key Findings}

In summary, key findings for RQs one and two are:

\begin{itemize}[leftmargin=*]
    \item \textbf{RQ1:} Users found the chatbots that explicitly referenced their previous utterances to be more engaging and intelligent.
    \item \textbf{RQ2:} Verbatim participants perceived the chatbot as collecting too much personal data compared to None participants. Reasoning for this can be found in qualitative findings (Section \ref{sec:interviews}).
\end{itemize}
\fi

\section{Discussion}

Here we discuss the implications of our study. We aimed to investigate the impact of a chatbot's format when referencing a user's utterances from a previous chatting session. By comparing high-level non-explicit references, verbatim references, and paraphrased references, we wanted to investigate effects on both positive user perceptions and privacy-related perceptions.
Our findings provide some empirical evidence that users value Verbatim and Paraphrase as more engaging and intelligent. However, (in support of Personalisation-Privacy Paradox \cite{awadpersonalization}) there is some evidence Verbatim and Paraphrase raised privacy concerns among users.

Although we did not find measurable differences in response quality between conditions, results indicated that people receiving non-explicit or verbatim references may be hesitant in providing their personal information.
Specifically, Verbatim participants were more concerned about the quantity of personal information being collected, and our interviews found that Verbatim participants raised concerns that the referencing style was ``unnerving'' and ``creepy''. 
Some None participants were hesitant providing complex utterances (as they doubted that the chatbot could understand them).
These findings could reflect the expectations of users before interacting with the chatbot \cite{hartmann2008framing,klaaren1994role,raita2011too,wilkinson2021or}. In order to abate these concerns, more clear consent could be sought and explanation of privacy practices could be provided \cite{seymour2023ignorance,lau2018alexa,phinnemore2023creepy} before using different referencing formats, and the abilities of the chatbot could be more clearly advertised to avoid user disappointment \cite{khadpe2020conceptual}.
%we did find that verbatim references raised more privacy concerns 

%different types of people
%Some see chatbot as tool (prefer verbatim) and some see more as conversational partner (prefer paraphrase).
%There are different types of people who may see the chatbot as either a tool or a conversation partner. 
Interviewees saw chatbots along a spectrum as either more of a conversation partner, or more of a tool to be used. 
%Those who view chatbots as more of a tool may prefer a chatbot that references them verbatim, while contrarily, those who view chatbots as conversation partners may prefer paraphrased responses.
Implications from this are that those who view chatbots as conversation partners may prefer paraphrased references, while contrarily, those who view chatbots as more of a tool may prefer a chatbot that references them verbatim.
Similarly, those with more faith in their own word compared to a chatbot (or no belief in chatbot intelligence or emotions) may prefer a verbatim reference format. 
This could be taken further by investigating the role of personality in user preferences for referencing formats. For example, users who are more extroverted or agreeable may prefer a more conversational (paraphrased) format, while users who are more introverted or conscientious may prefer a more direct and factual (verbatim) format.

Our findings also indicate the contextual nature of reference format. 
For example, if the user's utterance is akin to a ``contract'' to themselves (such as a goal for a healthy behaviour), they may want to see their utterance in its entirety in order to solidify their commitment. Similarly, if there is purpose in the user revisiting and developing on previous utterances (such as for creativity tasks or goal-setting) users may prefer their words to remain unchanged so as to build on their previous interaction. Equally, certain use cases (such as in legal settings) may require chatbots to be more conservative in their use of paraphrasing, or to provide verbatim quotes alongside the chatbot's paraphrase (akin to the use of mixed quotations in linguistics literature \cite{wilson_sperber_2000,Cappelen1997-CAPVOQ}).

%The study could be extended to investigate the impact of referencing formats on user behaviour. For example, do users who receive verbatim references tend to be more engaged with the chatbot, or do they find it more intrusive? Do paraphrased references lead to more successful goal attainment, or do users feel that the chatbot is not taking them seriously?

%People may prefer verbatim quotes for some of the reasons laid out in Section \ref{sec:referencing_recall}. 

This implies that chatbots could, in some cases, use a mixture of paraphrased and verbatim reference formats, depending on the content of the user's utterance. 
In the case of dental flossing, the chatbot could use paraphrased responses to reference a user's previous behaviour (flossing frequency), but maintain the user's utterance when referencing the user's behaviour strategy that they devised in the previous chatting session.
%Depending on the scenario, a chatbot could reference using paraphrase or verbatim formats.
%Based on length of a user’s utterances could change format.
%Based on the content of utterance.

Study findings also have implications for the design of chatbot interfaces. 
If chatbots are designed to reference utterances (e.g., verbatim quotes), designers need to be transparent to users, and ensure user control over their data and that user privacy is protected.
Similarly, if paraphrased references are used, the chatbot needs to ensure that the meaning of the user's original utterance is retained and that users do not feel that their utterance has been distorted.
%If chatbots are designed to provide verbatim references, they need to ensure that users have control over their data and that their privacy is protected. Similarly, if paraphrased references are used, the chatbot needs to ensure that the meaning of the user's original utterance is retained and that users do not feel that their utterance has been distorted.

%For more mundane utterances, it may not matter as greatly.

%Interesting as large language models become more conversational reinforcing the belief that chatbots should be more human-like....

%The study's findings may also have implications for the training of chatbots. If verbatim references are seen as more engaging, then chatbots may need to be trained to recognize and respond to specific keywords or phrases in user utterances. Similarly, if paraphrased references are seen as more respectful of privacy, chatbots may need to be trained to paraphrase user utterances in a way that accurately captures their meaning without revealing personal details.

%Finally, the study's findings could be applied to other areas of human-computer interaction beyond chatbots. For example, the design of automated email response systems or virtual assistants could be influenced by the study's findings, as both of these systems involve referencing previous user messages.

\section{Limitations and Future Work}

The user study was conducted over 3 weeks with one chatting session per week, which was not long enough to potentially encourage health behaviour change among participants.
%Additionally, the chatting sessions all took place on desktop/laptop devices, which (while ensuring consistency of input device) may prove less natural. 
%However, while one of the most popular mediums of communication is mobile phones for person-person communication, it is unclear if the rise of large language models (such as ChatGPT) would have normalised the use of desktops when communicating with a chatbot.
%Although historically one of the most popular mediums of communication is mobile phones for person-person communication, it is unclear if the rise of large language models (such as ChatGPT) would have normalised this consideration somewhat for participants.
Furthermore, we cannot claim generality over different chatbot referencing formats \cite{zhang2018making,el2021automatic}, sensitivity and intimacy of user data in references \cite{gomez2023sensitive}, domain of conversations with the chatbot, and input modalities.

Further work could investigate the use of referencing formats across different modalities. For example, while a voice-user interface (VUI) could also reference users verbatim or via paraphrasing, verbatim references could have the added dimension of using the voice of either an agent or of the user themselves \cite{NewsApple}. The added dimension of voice playback could raise addition concerns among users.
Additionally, alternative referencing formats (such as summarisation styles \cite{zhang2018making,el2021automatic,di2014hybrid}, or use of mixed quotations \cite{wilson_sperber_2000,Cappelen1997-CAPVOQ}) could be investigated. Choice of these could depend on factors such as the length, quantity, temporal spacing and content of utterances.
For example, for longer utterances, showing the entire utterance verbatim may prove unwieldy, adding to user burden \cite{eklundh1994use,severinson2010quote}.
%The use of a binary verbatim or paraphrase could be expanded further to include different types of paraphrasing (or summarisation \cite{zhang2018making,el2021automatic}) depending on the length, quantity and temporal spacing of user utterances.
%For example, for longer user utterances, showing the entire utterance verbatim may prove unwieldy.

\section{Conclusion}
This study investigates how the format used when a chatbot references user utterances from a previous chatting session affects a user's positive perceptions (chatbot intelligence and engagement) and privacy related perceptions. 
%This study investigates how the format of a chatbot's references to user utterances from a previous chatting session affects a user's positive perceptions (chatbot intelligence and engagement) and privacy related perceptions. 
%Our findings suggest that chatbots referencing  user utterances
Our findings suggest that if a chatbot references previous user utterances, both verbatim or by using paraphrases, it can lead to increased feelings of chatbot intelligence and engagement. 
Despite this, referencing user utterances can also raise privacy concerns among users. 
Our semi-structured interviews then investigated \textit{why} people have these privacy concerns. 
We discussed the implications of our findings for chatbot designers and researchers, and we provided recommendations
for the choice of referencing format.

%This study investigates how the language formality of a chatbot’s utterances affect the quality of self-disclosure from a user. We reported the results of two user studies conducted over AMT. Our findings suggest a formal conversational style may be perceived more positively when a chatbot is requesting sensitive health information, and may elicit more high quality user utterances when discussing a user’s health behaviour

\begin{acks}
This research is part of the programme DesCartes and is supported by the National Research Foundation, Prime Minister’s Office, Singapore under its Campus for Research Excellence and Technological Enterprise (CREATE) programme.
\end{acks}

%%
%% The next two lines define the bibliography style to be used, and
%% the bibliography file.
\bibliographystyle{ACM-Reference-Format}
\bibliography{sample-base}

%%
%% If your work has an appendix, this is the place to put it.
%\appendix

%\input{appendix.tex}

\end{document}